\documentclass[10pt,twocolumn,aps,pra,amsmath,amssymb,showpacs]{revtex4-1}
\usepackage{bm}
\usepackage{mathrsfs}
\usepackage{graphicx}
\usepackage[usenames,dvipsnames,svgnames,table]{xcolor}
\usepackage[unicode=true,
            pdfusetitle, 
            bookmarks=true,
            bookmarksnumbered=false,
            bookmarksopen=false,
            breaklinks=true,
            pdfborder={0 0 0},
            backref=false,
            colorlinks=true]{hyperref}
\hypersetup{linkcolor=NavyBlue,urlcolor=NavyBlue,citecolor=NavyBlue}

\usepackage{amsthm}

\providecommand{\propositionname}{Proposition}

\usepackage{tikz-cd}

\newcommand{\cH}{\mathcal{H}}

\begin{document}

\title{When the assignment map is completely positive}
\author{Iman Sargolzahi}
\email{sargolzahi@neyshabur.ac.ir; sargolzahi@gmail.com}
\affiliation{Department of Physics, University of Neyshabur, Neyshabur, Iran}
\author{Sayyed Yahya Mirafzali}
\email{y.mirafzali@vru.ac.ir}
\affiliation{Department of Physics, Faculty of Science, Vali-e-Asr University of Rafsanjan, Rafsanjan, Iran}

\begin{abstract}
Finding the general set of system-environment states $\lbrace\rho_{SE}\rbrace $ for which the reduced dynamics of the system is completely positive (CP) is the subject of some recent works. An advance in this context appeared in  [X.-M. Lu, \href{http://dx.doi.org/10.1103/PhysRevA.93.042332}{Phys. Rev. A {\bf 93}, 042332 (2016)}], where the problem was solved for the case of CP assignment map. Here, we restate this result using the framework introduced in  [J. M. Dominy \textit{et al.}, \href{http://link.springer.com/article/10.1007/s11128-015-1148-0}{ Quantum Inf. Process. {\bf 15}, 465 (2016)}]. This, we think,  clarifies the mentioned result better and so leads to a generalization of it, straightforwardly.

\end{abstract}


\maketitle


\section{Introduction}
Consider a closed finite-dimensional quantum system which evolves according
 to
\begin{equation}
\label{eq:one}
\begin{aligned}
\rho\rightarrow\rho^{\prime} =Ad_{U}(\rho)\equiv U\rho U^{\dagger},
\end{aligned}
\end{equation}
where $\rho$ and $\rho^{\prime}$ are the initial and final states (density operators) of the system, respectively, and $U$ is a unitary operator ($UU^{\dagger}= U^{\dagger}U=I$, where $I$ is the identity operator).

In general, the system is not closed and interacts with its environment. We can consider the entire system-environment as a closed quantum system which evolves as Eq. (\ref{eq:one}). So the reduced state of the system after the evolution is given by
\begin{equation}
\label{eq:two}
\begin{aligned}
\rho_{S}^{\prime}=\mathrm{Tr_{E}} \circ Ad_{U}(\rho_{SE})=\mathrm{Tr_{E}}\left( U \rho_{SE}U^{\dagger}\right), 
\end{aligned}
\end{equation} 
where $\rho_{SE}$ is the initial state of the combined system-environment quantum system and $U$ acts on the whole Hilbert space of system-environment. Now a natural  question is  what is the relation between the initial state of the system $\rho_{S}=\mathrm{Tr_{E}}(\rho_{SE})$ and its final state $\rho_{S}^{\prime}$? Can this relation be represented by a map and - if so -  what kind of map? 

Consider the case that the set of possible initial states of system-environment is factorized: 
$ \mathcal{S}=\lbrace\rho_{SE}=\rho_{S}\otimes\tilde{\omega}_{E}\rbrace $,
where $\rho_{S}$ are arbitrary states of the system, but $\tilde{\omega}_{E}$ is a fixed state of environment. It is famous that for this special case the reduced dynamics of the system is given by a completely positive (CP) map \cite{1}. But, as we will see in the next section, it is not so in general. Therefore, finding the general set of initial $ \rho_{SE} $ for which the reduced dynamics of the system is CP, has become the subject of some recent studies \cite{11, 12, 13, 14, 15, 16}, which we will review them in Sec.~\ref{sec:CP reduced_dynamics}.

In this context, the most general set of initial $ \rho_{SE} $  known prior to our present work (that leads to CP reduced dynamics)  has been introduced in Ref. \cite{ 15}. In fact, as proven in Ref. \cite{16},  this is the final possible generalization, if we restrict ourselves to the case of CP assignment map.

In this paper, we will restate the result of Ref. \cite{16}, using the framework introduced in Ref.  \cite{3}. This will help to clarify the result of Ref.  \cite{16} better. Specially, our treatment will highlight the condition of $U$-consistency, for arbitrary $U$, which is needed to achieve the result of Ref. \cite{16}. We will give the details in Sec.~\ref{sec:assignment map}.

In  Sec.~\ref{sec:U-consistent}, we will generalize the result of Refs. \cite{ 15, 16}; i.e. we will find a more general set of initial $ \rho_{SE} $ which leads to CP reduced dynamics. So, it includes all the previous results in this context. However, this generalization is rather straightforward, using the framework of Ref. \cite{3}. 

We will end this paper in  Sec.~\ref{sec:summary}, with a summary of our results.

\section{Reduced dynamics of  open quantum system}\label{sec:reduced_dynamics}

There was a tendency to assume the CP maps as the only possible quantum dynamics of a system. But, using Eq. (\ref{eq:two}), it can be shown simply that it is not so for open quantum systems.
In fact, the evolution $\rho_{S}\rightarrow\rho_{S}^{\prime}$ may not be represented by a map, in general \cite{2}. This can be illustrated by the following simple example \cite{3}. Assume that the initial state of the system-environment can be chosen from the set
 $\mathcal{S}=\lbrace{\rho_{S}^{(1)}\otimes\rho_{E}^{(1)},\;\rho_{S}^{(1)}\otimes\rho_{E}^{(2)},\;\cdots}\rbrace$ and the evolution of the system-environment is given by the swap operator $U_{sw}\vert\psi\rangle\vert\phi\rangle=\vert\phi\rangle\vert\psi\rangle$. Now for the case that the initial state of the system is $\rho_{S}^{(1)}$, there are (at least) two possible final states:
 \begin{equation*}
\begin{aligned}
\rho_{S}^{(1)\,\prime}=\mathrm{Tr_{E}}\left( U_{sw}\,\rho_{S}^{(1)}\otimes\rho_{E}^{(1)}\,U_{sw}^{\dagger}\right)=\rho_{E}^{(1)}, \\  \rho_{S}^{(2)\,\prime}=\mathrm{Tr_{E}}\left( U_{sw}\,\rho_{S}^{(1)}\otimes\rho_{E}^{(2)}\,U_{sw}^{\dagger}\right)=\rho_{E}^{(2)}. 
\end{aligned}
\end{equation*}
So the evolution from $\rho_{S}$ to $\rho_{S}^{\prime}$ cannot be represented by a map (a map, by definition, assigns to each initial state, e.g. $\rho_{S}^{(1)}$, only one final state).
 
Even if the evolution from $\rho_{S}$ to $\rho_{S}^{\prime}$ can be represented by a map, this map is not linear, in general \cite{4}. This also can be represented by the following simple example \cite{5}. Consider an arbitrary non-linear map $\rho_{S}\rightarrow\Phi(\rho_{S})$. Assume that the set of passible initial $\rho_{SE}$ is $\mathcal{S}=\lbrace{\rho_{S}\otimes\Phi(\rho_{S})\rbrace}$ and the evolution of the system and environment is again given by the swap operator. So $\mathrm{Tr_{E}}\left( U_{sw}\,\rho_{S}\otimes\Phi(\rho_{S})\,U_{sw}^{\dagger}\right)=\Phi(\rho_{S})$, which is, by assumption, a non-linear function of the initial state $\rho_{S}$. Therefore, in order that Eq. (\ref{eq:two}) leads to a linear map from $\rho_{S}$ to $\rho_{S}^{\prime}$, there must be some restrictions on the set of possible initial states $\rho_{SE}$ or on the possible evolution $U$ \cite{3, 6}.

However, if the reduced dynamics of the system from $\rho_{S}$ to $\rho_{S}^{\prime}$ can be given by a linear map $\Psi$, then it can be shown readily that this $\Psi$ is Hermitian, i.e. maps each Hermitian operator to a Hermitian operator (details are given in Sec.~\ref{sec:assignment map}).
Now, an important result is that for each linear trace-preserving Hermitian map from $\rho_{S}$ to $\rho_{S}^{\prime}$, there exists an operator sum representation  in the following form:
\begin{equation}
\label{eq:three}
\begin{aligned}
\rho_{S}^{\prime}=\sum_{i}e_{i}\,\tilde{E_{i}}\,\rho_{S}\,\tilde{E_{i}}^{\dagger},\ \ \ \sum_{i}e_{i}\,\tilde{E_{i}}^{\dagger}\tilde{E_{i}}=I_{S},
\end{aligned}
\end{equation}
where $\tilde{E_{i}}$ are linear operators and $e_{i}$ are real coefficients \cite{8, 7, 3}. Note that the relation $\sum_{i}e_{i}\,\tilde{E_{i}}^{\dagger}\tilde{E_{i}}=I_{S}$ comes from the fact that the map is trace-preserving, as expected from Eq. (\ref{eq:two}). It is also worth noting that $\tilde{E_{i}}$ and $e_{i}$ are fixed, i.e. they are independent of initial state $\rho_{S}$.

If all of the coefficients $e_{i}$ in Eq. (\ref{eq:three}) are positive, then we call the map completely positive and rewrite Eq. (\ref{eq:three}) in the following form:  
\begin{equation}
\label{eq:four}
\begin{aligned}
\rho_{S}^{\prime}=\sum_{i}E_{i}\,\rho_{S}\,E_{i}^{\dagger},\ \ \ \sum_{i}E_{i}^{\dagger}E_{i}=I_{S},
\end{aligned}
\end{equation}
where $E_{i}\equiv\sqrt{e_{i}}\,\tilde{E_{i}}$.  In the languge of Ref.~\cite{3}, we should call a map which is given by Eq.~(\ref{eq:four}), a completely positively trace-preserving extensible map. However, in this paper, we simply refer to such a map a completely positive (CP) map.

 The simplest standard example which leads to CP reducd dynamics is when the initial state of the system-environment is $\rho_{SE}=\rho_{S}\otimes\tilde{\omega}_{E}$, where $\rho_{S}$ is an arbitrary state of the system, but $\tilde{\omega}_{E}$ is a fixed state of environment. For this case, it can be shown simply that Eq. (\ref{eq:two}), for an arbitrary $U$, gives us a CP map \cite{1}

Besides the factorized initial states $\rho_{S}\otimes\tilde{\omega}_{E}$, there exist other sets of initial states $\rho_{SE}$ for which the reduced dynamics is CP  for arbitrary $U$. This is the subject of some recent works \cite{11, 12, 13, 14, 15, 16}, which we will review them in the next section.

\section{Sets of initial states which lead to CP reduced dynamics}\label{sec:CP reduced_dynamics}

As we have seen in the previous section, the reduced dynamics of an open quantum system is not given by a CP map, in general. In fact, the CP-ness of the reduced dynamics has been proven only for some restricted sets of initial $ \rho_{SE} $, which we will review them in the following.
   
The simplest case which leads to the CP reduced dynamics is when the initial state of the system-environment is factorized, i.e. when the set of possible initial $\rho_{SE}$ is
\begin{equation}
 \label{eq:five-a}
\begin{aligned}
\mathcal{S}=\lbrace\rho_{SE}=\rho_{S}\otimes\tilde{\omega}_{E}\rbrace ,
\end{aligned}
\end{equation}
where $\rho_{S}$ are arbitrary states of the system, but $\tilde{\omega}_{E}$ is a fixed state of environment.
 Let the eigen-decomposition of $\tilde{\omega}_{E}$ be as $\tilde{\omega}_{E}=\sum_{l}\lambda_{l}\;\vert \mu_{E}^{(l)}\rangle\langle \mu_{E}^{(l)}\vert$  ($\lambda_{l}\geq 0$, $\sum_{l}\lambda_{l}=1$). So, for an arbitrary $U$ in  Eq. (\ref{eq:two}), we have
\begin{equation}
\label{eq:five}
\begin{aligned}
\rho^{\prime}_{S} =\mathrm{Tr_{E}}\left( U\rho_{S}\otimes\tilde{\omega}_{E} U^{\dagger}\right)  \quad\qquad \;\qquad\\ =\sum_{kl}\lambda_{l}\;\langle k_{E}\vert U\vert \mu_{E}^{(l)}\rangle\;\,\rho_{S}\;\,\langle \mu_{E}^{(l)}\vert U^{\dagger}\vert k_{E}\rangle \quad\;\;\\
=\sum_{kl}E_{kl}\;\rho_{S}\; E_{kl}^{\dagger}, \;\;\; \sum_{kl}E_{kl}^{\dagger}E_{kl}=I_{S}, 
\end{aligned}
\end{equation}
where $ \lbrace\vert k_{E}\rangle\rbrace$ is an orthonormal basis of the Hilbert space of the environment $\cH_E$ and $E_{kl}\equiv\sqrt{\lambda_{l}}\;\langle k_{E}\vert U\vert \mu_{E}^{(l)}\rangle$ are linear operators on the Hilbert space of the system $\cH_S$. In addition, $ I_{S} $ is the identity operator on $\cH_S$.  Therefore, the dynamics of the factorized initial $\rho_{SE}$ always reduces to a CP map.

 In addition to the factorized initial states , one can find other possible initial sets which also yield CP reduced dynamics. The first, introduced in Ref. \cite{11}, is
\begin{equation}
\label{eq:six}
\begin{aligned}
\mathcal{S}=\lbrace\rho_{SE}=\sum_{i}p_{i}\vert\tilde{i}_{S}\rangle\langle\tilde{i}_{S}\vert\otimes\tilde{\omega}_{i}\rbrace ,
\end{aligned}
\end{equation}
where $\lbrace p_{i}\rbrace$ is arbitrary probability distribution ($p_{i}\geq 0$, $\sum_{i}p_{i}=1$), but $\lbrace\vert\tilde{i}_{S}\rangle\rbrace$ is a fixed orthonormal basis for $\cH_S$ and $\tilde{\omega}_{i}$ are fixed density operators on $\cH_E$. Let $\tilde{\Pi}_{i}=\vert\tilde{i}_{S}\rangle\langle\tilde{i}_{S}\vert$ and the eigen-decomposition of $\tilde{\omega}_{i}$ be as $\tilde{\omega}_{i}=\sum_{l}\lambda_{il}\;\vert \mu_{E}^{(il)}\rangle\langle \mu_{E}^{(il)}\vert$. 
 Now, for arbitrary $U$ in Eq. (\ref{eq:two}), we have 
\begin{equation}
\label{eq:seven}
\begin{aligned}
\rho_{S}^{\prime}=\mathrm{Tr_{E}}\left( U\,\left(\sum_{i}p_{i}\,\tilde{\Pi}_{i}\otimes\tilde{\omega}_{i}\right)\,U^{\dagger}\right) \qquad\quad \\ 
=\sum_{ikl} p_{i}\lambda_{il}\,\langle k_{E}\vert U\vert \mu_{E}^{(il)}\rangle\;\tilde{\Pi}_{i}\,\langle \mu_{E}^{(il)}\vert U^{\dagger}\vert k_{E}\rangle\,\qquad\;\; \\
=\sum_{ikl}\lambda_{il}\,\langle k_{E}\vert U\vert \mu_{E}^{(il)}\rangle\;\tilde{\Pi}_{i}\,\rho_{S}\,\tilde{\Pi}_{i}\,\langle \mu_{E}^{(il)}\vert U^{\dagger}\vert k_{E}\rangle\,\quad \\
=\sum_{ikl}E_{ikl}\,\rho_{S}E_{ikl}^{\dagger},\; \quad
  \sum_{j}E_{ikl}^{\dagger}\,E_{ikl}=I_{S},\; \ \ 
\end{aligned}
\end{equation}
where $E_{ikl}\equiv D_{ikl}\,\tilde{\Pi}_{i}$ and $D_{ikl}\equiv \sqrt{\lambda_{il}}\,\langle k_{E}\vert U\vert \mu_{E}^{(il)}\rangle$. As we see, Eq. (\ref{eq:seven}) is in the form of Eq. (\ref{eq:four}); so it is a CP map.

It is also worth noting that, in addition to the above CP map in Eq. (\ref{eq:seven}), one can find other CP maps which equivalently  describe the reduced dynamics of the system. This is due to the fact that one can find more than one CP \textit{assignment maps} $\Lambda$ such that their effects on all $\rho_{S}\in   \mathrm{Tr_{E}}\mathcal{S}$ are the same \cite{77}. (The assignment map is a map which assigns to  each $\rho_{S}\in   \mathrm{Tr_{E}}\mathcal{S}$  a $\rho_{SE}\in  \mathcal{S}$ such that $ \mathrm{Tr_{E}}(\rho_{SE})=\rho_{S}$.) Obviously, the effects of these different assignment maps $\Lambda$  on (some of the) states $\rho_{S}\notin  \mathrm{Tr_{E}}\mathcal{S}$ are different.

The result of Ref. \cite{11} is then extended in Refs. \cite{12, 14} where it has been shown that the set of initial states 
\begin{equation}
\label{eq:eight}
\begin{aligned}
\mathcal{S}=\lbrace\rho_{SE}=\bigoplus_{i=1}^{m}p_{i}\rho_{S}^{(i)}\otimes\tilde{\omega}_{i}\rbrace ,\quad\\
\cH_S=\cH_S^{(1)}\oplus\cH_S^{(2)}\oplus\cdots\oplus\cH_S^{(m)},\;
\end{aligned}
\end{equation}
also yields CP reduced dynamics, for arbitrary $U$ in Eq. (\ref{eq:two}). In Eq. (\ref{eq:eight}), $\lbrace p_{i}\rbrace$ is arbitrary probability distribution, $\rho_{S}^{(i)}$ is an arbitrary state on $\cH_S^{(i)}$, but $\tilde{\omega}_{i}$ are fixed states on $\cH_E$.

The next generalization was given in Refs. \cite{13,14}, where the CP reduced dynamics was proven for the following initial states $\rho_{SE}$ and arbitrary $U$ in Eq. (\ref{eq:two}):    
\begin{equation}
\label{eq:nine}
\begin{aligned}
\mathcal{S}=\lbrace\rho_{SE}=\bigoplus_{i=1}^{m^{\prime}}p_{i}\,\tilde{\omega}_{SE}^{(i)}+\bigoplus_{i=m^{\prime}+1}^{m}p_{i}\rho_{S}^{(i)}\otimes\tilde{\omega}_{i}\rbrace ,\quad\\
\cH_S=\cH_S^{(1)}\oplus\cH_S^{(2)}\oplus\cdots\oplus\cH_S^{(m)}.\;\qquad\quad
\end{aligned}
\end{equation}
Again, $\lbrace p_{i}\rbrace$ is arbitrary probability distribution and $\rho_{S}^{(i)}$ is arbitrary state on $\cH_S^{(i)}$, but $\tilde{\omega}_{SE}^{(i)}$ is a fixed state on $\cH_S^{(i)}\otimes\cH_E$ and $\tilde{\omega}_{i}$ are fixed states on $\cH_E$. The operator sum representations for the CP reduced dynamics, given by the initial states $\rho_{SE}$ in Eqs. (\ref{eq:eight}) and (\ref{eq:nine}), are given in Ref. \cite{14}.

The final generalization (prior to our present work) is that of Ref. \cite{15}, which we will write it in the  form introduced in Ref. \cite{16}. There, it has been shown that if the set of initial states $\rho_{SE}$ is given by
\begin{equation}
\label{eq:ten}
\begin{aligned}
\mathcal{S}=\lbrace\rho_{SE}=\bigoplus_{i}p_{i}\,\rho_{L_{i}}\otimes\tilde{\omega}_{R_{i}E}\rbrace ,\quad\\
\cH_S=\bigoplus_{i}\cH_{L_{i}}\otimes\cH_{R_{i}},\;\qquad\quad
\end{aligned}
\end{equation}
then the reduced dynamics of the system, given by Eq. (\ref{eq:two}) with arbitrary $U$, is a CP map. In the above equation, $\lbrace p_{i}\rbrace$ is arbitrary probability distribution, $\rho_{L_{i}}$ is arbitrary state on $\cH_{L_{i}}$, but $\tilde{\omega}_{R_{i}E}$ is a fixed state on $\cH_{R_{i}}\otimes\cH_E$.

It can be shown simply that all the previous sets, given in Eqs. (\ref{eq:five-a}), (\ref{eq:six}), (\ref{eq:eight}) and (\ref{eq:nine}), are special cases of Eq. (\ref{eq:ten}) \cite{16}. For example, the factorized initial state $\rho_{SE}=\rho_{S}\otimes\tilde{\omega}_{E}$ is due to the case that the summation in Eq. (\ref{eq:ten}) includes only one term; so $\cH_{S}=\cH_{L}\otimes\cH_R$, where $\cH_{R}$ is a trivial one-dimensional Hilbert space. In addition, the operator sum representation for the CP reduced dynamics of the system, with the initial $\rho_{SE}\in \mathcal{S}$ in Eq. (\ref{eq:ten}) and arbitrary $U$ in Eq.(\ref{eq:two}), is given in Ref. \cite{16}.

Now an important result, proven in Ref. \cite{16}, is that Eq. (\ref{eq:ten}) is in fact the final possible generalization, if we restrict ourselves to the case of CP assignment map.   In other words, the set of initial $\rho_{SE}$ is given by Eq. (\ref{eq:ten}), if and only if, the assignment map is a CP map. In the next section, we restate this result, using the framework introduced in Ref. \cite{3}, which will help to clarify this result better and allow us to generalize it in Sec.~\ref{sec:U-consistent}. 

Let us end this section with an additional remark. The set $\mathcal{S}$, in Eq. (\ref{eq:ten}), can be written as the \textit{steered} set from a fixed tripartite state $\tilde{\omega}_{ASE}$. We define
\begin{equation}
\label{eq:ten-a}
\begin{aligned}
\tilde{\omega}_{ASE}=\bigoplus_{i}\tilde{q}_{i}\,\tilde{\omega}_{AL_{i}}\otimes\tilde{\omega}_{R_{i}E},
\end{aligned}
\end{equation}
on the Hilbert space $\cH_{A}\otimes\cH_S\otimes\cH_E$, where $\cH_{A}$ is an ancillary Hilbert space, 
$\lbrace \tilde{q}_{i}\rbrace$ is a probability distribution, $\tilde{\omega}_{AL_{i}}$ is a state on $\cH_{A}\otimes\cH_{L_{i}}$, and  $\tilde{\omega}_{R_{i}E}$ are those states introduced in Eq. (\ref{eq:ten}),  on $\cH_{R_{i}}\otimes\cH_E$.
The set of steered states, from performing  measurements on the part $A$ of  $\tilde{\omega}_{ASE}$, is \cite{ 15, 16}:
\begin{equation}
\label{eq:1a}
\begin{aligned}
\mathcal{S}=\left\lbrace \rho_{SE}= \frac{\mathrm{Tr_{A}}[(P_{A}\otimes I_{SE})\tilde{\omega}_{ASE}]}{\mathrm{Tr}[(P_{A}\otimes I_{SE})\tilde{\omega}_{ASE}]} ,  P_{A}>0 \right\rbrace ,  
\end{aligned}
\end{equation}
where $P_{A}$ is  arbitrary positive operator on $\cH_{A}$ such that $\mathrm{Tr}[(P_{A}\otimes I_{SE})\tilde{\omega}_{ASE}]>0$ and $I_{SE}$ is the identity operator on $\cH_{S}\otimes\cH_{E}$. Note that, up to a positive  factor, $P_{A}$ can be considered as an element of a POVM. Now, by an appropriate choice of the states $\tilde{\omega}_{AL_{i}}$ in Eq. (\ref{eq:ten-a}), it can be shown
 that the steered set $\mathcal{S}$ in Eq. (\ref{eq:1a}) coincides with that of Eq. (\ref{eq:ten}) \cite{16}.

A tripartite state which can be decomposed as Eq. (\ref{eq:ten-a}) is called a \textit{Markov} state \cite{10}. Therefore, for the steered set from a Markov state $\tilde{\omega}_{ASE}$, the reduced dynamics of the system, for arbitrary $U$ in Eq.  (\ref{eq:two}),  is CP. Interestingly, the reverse is also true: If for a steered set, from a fixed tripartite state $\tilde{\omega}_{ASE}$, the reduced dynamics of the system, for arbitrary $U$, is CP, then  $\tilde{\omega}_{ASE}$ is a Markov state \cite{15}.

A Markov state $\tilde{\omega}_{ASE}$ possesses another interesting property too: For arbitrary unitary evolution of the system-environment $U_{SE}$, $\tilde{\omega}_{ASE}^{\prime}=id_{A}\otimes Ad_{U_{SE}}(\tilde{\omega}_{ASE})$ where $id_{A}$ is the identity map on $A$, the following \textit{quantum data processing inequality} is satisfied  \cite{15}:
\begin{equation}
\label{eq:15}
I(A:S)_{\tilde{\omega}_{AS}}\geq I(A:S)_{\tilde{\omega}_{AS}^{\prime}},
\end{equation} 
where $\tilde{\omega}_{AS}=\mathrm{Tr}_{E}(\tilde{\omega}_{ASE})$ and $\tilde{\omega}_{AS}^{\prime}=\mathrm{Tr}_{E}(\tilde{\omega}_{ASE}^{\prime})$. In the above equation, $I(A:S)$ is the \textit{mutual information}. For a bipartite state $\omega_{AS}$,  the mutual information  is defined as $I(A:S)_{\omega}=S(\omega_A)+S(\omega_S)-S(\omega_{AS})$, where $\omega_{A}=\mathrm{Tr_{S}}(\omega_{AS})$ and $\omega_{S}=\mathrm{Tr_{A}}(\omega_{AS})$ and $S(\omega)$ is the von Neumann entropy of the state $\omega$: $S(\omega)=- \mathrm{Tr}(\omega\log\omega)$ \cite{1}. 
Interestingly, the reverse is also true: If, for a tripartite state $\tilde{\omega}_{ASE}$, the inequality  (\ref{eq:15}) is satisfied for arbitrary $U_{SE}$, then $\tilde{\omega}_{ASE}$ is a Markov state as Eq. (\ref{eq:ten-a})  \cite{15}.

In summary, we have seen that for the steered set $\mathcal{S}$, as  Eq. (\ref{eq:1a}),
 the reduced dynamics of the system is CP, for arbitrary $U_{SE}$, if and only if, the $\tilde{\omega}_{ASE}$ is a Markov state. On the other hand, the $\tilde{\omega}_{ASE}$ is a Markov state, if and only if, the   
quantum data processing inequality  (\ref{eq:15}) is satisfied for arbitrary $U_{SE}$. Therefore, for the steered set $\mathcal{S}$ in Eq. (\ref{eq:1a}), the CP-ness of the reduced dynamics of the system, for arbitrary $U_{SE}$, is equivalent to the satisfaction of the quantum data processing inequality  (\ref{eq:15}), for  arbitrary $U_{SE}$ \cite{15}.

\section{When the assignment map is CP} \label{sec:assignment map}
As mentioned in Sec.~\ref{sec:reduced_dynamics}, if the reduced dynamics of the system in  Eq. (\ref {eq:two}) can be represented by a linear map $\Psi$, then it can be shown simply that this $\Psi$ is Hermitian. 
  Consider the set $\mathcal{S}_{S}$ of initial $\rho_{S}=\mathrm{Tr_{E}}(\rho_{SE})$ for which $\rho_{S}^{\prime}=\mathrm{Tr_{E}} \circ Ad_{U}(\rho_{SE})$ are given by $\Psi(\rho_{S})$: $\rho_{S}^{\prime}= \Psi(\rho_{S})$. $\mathcal{S}_{S}$ is called the \textit{physical domain} \cite{3} or the \textit{compatibility domain}  \cite{8} of $\Psi$. Since the Hilbert space of the system $\cH_S$ is finite-dimensional, one can find a set $\mathcal{S}_{S}^{\prime}\subset \mathcal{S}_{S}$ including a finite number of $\rho_{S}^{(i)}\in \mathcal{S}_{S}$ which are linearly independent and other states in $\mathcal{S}_{S}$ can be decomposed as linear combinations of them: $\mathcal{S}_{S}^{\prime}=\lbrace\rho_{S}^{(1)},\;\rho_{S}^{(2)},\;\cdots ,\;\rho_{S}^{(m)}\rbrace$, where $m$ is an integer and $m\leq(d_{S})^{2}$ ( $d_{S}$ is the dimension of $\cH_S$, so $(d_{S})^{2}$ is the dimension of $\mathcal{L}(\cH_S)$, the space of linear operators on $\cH_S$), and, for each $\rho_{S}\in \mathcal{S}_{S}$, we have $\rho_{S}=\sum_{i=1}^{m}a_{i}\,\rho_{S}^{(i)}$ with real $a_{i}$. Now any Hermitian operator $A$, which can be expanded by $\rho_{S}^{(i)}\in \mathcal{S}_{S}^{\prime}$, is obviously mapped by the linear map $\Psi$ to a Hermitian operator: $A=\sum_{i=1}^{m}c_{i}\,\rho_{S}^{(i)}$, with real $c_{i}$, so $\Psi(A)=\sum_{i=1}^{m}c_{i}\,\Psi(\rho_{S}^{(i)})$. $\Psi(\rho_{S}^{(i)})$ are density operators, therefore $\Psi(A)$ is Hermitian. In other words, $\Psi(A)$ is Hermitian for all Hermitian operators $A$ which can be expanded by $\rho_{S}^{(i)}\in \mathcal{S}_{S}^{\prime}$. Even if $m<(d_{S})^{2}$ one can easily extend $\Psi$ to  construct a linear Hermitian map on the whole $\mathcal{L}(\cH_S)$.
  
A general framework for linear trace-preserving Hermitian maps, arising from  Eq. (\ref{eq:two}), has been developed in Ref. \cite{3}. This framework can be used to prove interesting results. For example, in Ref. \cite{9}, it has been shown that the \textit{physically relevant} part of any Hermitian map $\Psi$, can  represent a possible reduced dynamics of the system. By physically relevant part of $\Psi$, we mean the restriction of $\Psi$ to those initial states $\rho_{S}$ which are mapped by $\Psi$ to density operators, i.e. $\Psi(\rho_{S})$ are also density operators. In the following, we use this framework to restate the main result of Ref. \cite{16}.

For the finite-dimensional Hilbert space $\cH_{S}\otimes\cH_E$, the set of linear operators on $\cH_{S}\otimes\cH_E$, i.e. $\mathcal{L}(\cH_{S}\otimes\cH_E)$, is also a finite-dimensional vector space.  Consider the set $\mathcal{S}\subseteq \mathcal{D}_{SE}$, where $\mathcal{D}_{SE}$ is the set of density operators on $\cH_{S}\otimes\cH_E$. Obviously $\mathcal{S}\subset \mathcal{L}(\cH_{S}\otimes\cH_E)$; so one can find a set $\mathcal{S}^{\prime}$, including a finite number of       
$\rho_{SE}^{(l)}$ in $\mathcal{S}$, which are linearly independent and decompose other elements of 
$\mathcal{S}$ as linear combinations of these $\rho_{SE}^{(l)}\in \mathcal{S}^{\prime}$. For example, if $\mathcal{S}^{\prime}=\lbrace\rho_{SE}^{(1)},\,\rho_{SE}^{(2)},\,\cdots,\,\rho_{SE}^{(n)}\rbrace$, we can write each member of $\mathcal{S}$ as $\rho_{SE}=\sum_{l=1}^{n} a_{l}\,\rho_{SE}^{(l)}$, where the real coefficients $a_{l}$ are unique.  Finally, let us define $\mathcal{V}\subseteq \mathcal{L}(\cH_{S}\otimes\cH_{E})$ as the subspace spanned by $\rho_{SE}^{(l)}\in \mathcal{S}^{\prime}$; i.e., for each $X\in \mathcal{V}$, we have $X=\sum_{l}c_{l}\,\rho_{SE}^{(l)}$ with unique complex coefficients $c_{l}$. Obviously $\mathcal{S}\subset\mathcal{V}$.

Now consider a linear map $\Psi$ on $\mathrm{Tr}_{E}\mathcal{V}$ in the form of Eq. (\ref{eq:two}). So, for each $x\in\mathrm{Tr}_{E}\mathcal{V}$, we have ($x=\mathrm{Tr}_{E}X$, $X\in \mathcal{V}$): 
\begin{equation}
\label{eq:eleven}
\begin{aligned}
x^{\prime}=\Psi(x)=\mathrm{Tr_{E}} \circ Ad_{U}(X)=\mathrm{Tr_{E}}\left( U\,X\,U^{\dagger}\right). \\   
\end{aligned}
\end{equation}
The first obvious requirement that such a map $\Psi$ can be defined, is the \textit{$U$-consistency} of the $\mathcal{V}$ \cite{3}; i.e. if for two operators $X_{1}$ and $X_{2}\in\mathcal{V}$, we have $\mathrm{Tr}_{E}X_{1}=\mathrm{Tr}_{E}X_{2}=x$, then we must have $\mathrm{Tr}_{E}\circ Ad_{U}(X_{1})=\mathrm{Tr}_{E}\circ Ad_{U}(X_{2})=\Psi(x)$. In Ref. \cite{3}, it has been shown that if $\mathcal{S}$  is convex and $U$-consistent, then $\mathcal{V}$ is $U$-consistent too. Now, one can define an  assignment map $\Lambda_{1}$ ($\Lambda_{2}$) in the following form: $\Lambda_{1}(x)=X_{1}$ ($\Lambda_{2}(x)=X_{2}$). So $\Psi$ in Eq. (\ref{eq:eleven}) can be written as $\Psi(x)=\mathrm{Tr}_{E}\circ Ad_{U}\circ\Lambda_{1}(x)$ ($\Psi(x)=\mathrm{Tr}_{E}\circ Ad_{U}\circ\Lambda_{2}(x)$).

It has been shown in Ref. \cite{3} that if $\mathcal{V}$ is \textit{U}-consistent for arbitrary unitary  $U\in\mathcal{L}(\cH_{S}\otimes\cH_{E})$, then for each $x\in\mathrm{Tr}_{E}\mathcal{V}$, there is only one $X\in\mathcal{V}$ for which we have $\mathrm{Tr}_{E}X=x$. So, there is only one way to define the assignment map $\Lambda$. 

The assignment map $\Lambda$ is Hermitian and so has an operator sum representation as Eq. (\ref{eq:three})  \cite{3}. Now let's restrict ourselves to the case that $\Lambda$ is in addition a CP map and so has an operator sum representation as Eq. (\ref{eq:four}). (The extension of) $\Lambda$ is a map form $\mathcal{L}(\cH_{S})$ to $\mathcal{L}(\cH_{S}\otimes\cH_{E})$. To make the input and output spaces the same, we redefine     $\Lambda$ in the following way: If for a $x\in\mathcal{L}(\cH_{S})$ we have $\Lambda(x)=X\in\mathcal{L}(\cH_{S}\otimes\cH_{E})$, we set $\Lambda(x\otimes\vert 0_{E}\rangle\langle 0_{E}\vert)=X$ where $\vert 0_{E}\rangle$ is a fixed state in $\cH_{E}$. This redefinition helps us to write $\Lambda$ in the following form. One can find an ancillary Hilbert space $\cH_{C}$, a fixed state $\vert 0_{C}\rangle\in\cH_{C}$ and a unitary operator $V$ on $\cH_{S}\otimes\cH_{E}\otimes\cH_{C}$ in such a way that the CP map $\Lambda$ can be written as \cite{1}:
\begin{equation}
\label{eq:twelve}
\begin{aligned}
\Lambda(x)=\Lambda(x\otimes\vert 0_{E}\rangle\langle 0_{E}\vert) \qquad\qquad\qquad\quad \;\;\\
=\mathrm{Tr_{C}}\left(V\,(x\otimes\vert 0_{E}\rangle\langle 0_{E}\vert\otimes\vert 0_{C}\rangle\langle 0_{C}\vert)\,V^{\dagger}\right).
\end{aligned}
\end{equation}

The next observation is based on the useful result of Ref. \cite{17}. Note that for arbitrary state $\rho_{S}\in\mathrm{Tr_{E}}\mathcal{V}$, we have 
\begin{equation}
\label{eq:thirteen}
\begin{aligned}
\hat{\Phi}(\rho_{S})\equiv\mathrm{Tr_{E}} \circ \Lambda(\rho_{S})=\rho_{S}.
\end{aligned}
\end{equation}   
$\hat{\Phi}$ is a CP map, since $\Lambda$ is a CP map by assumption and the partial trace $\mathrm{Tr_{E}}$ is also a CP map \cite{1}. In addition note that
\begin{equation}
\label{eq:fourteen}
\begin{aligned}
\hat{\Phi}(\rho_{S})
=\mathrm{Tr_{EC}}\left(V\,(\rho_{S}\otimes\vert 0_{E}\rangle\langle 0_{E}\vert\otimes\vert 0_{C}\rangle\langle 0_{C}\vert)\,V^{\dagger}\right),
\end{aligned}
\end{equation}
with the unitary $V$ introduced in Eq. (\ref{eq:twelve}). 

In Ref. \cite{17}, it has been shown that if for a set of states $\mathcal{S}_{S}\equiv\mathrm{Tr_{E}}(\mathcal{V})\cap\, \mathcal{D}_{S}$ (where $\mathcal{D}_{S}$ is the set of density operators on $\cH_S$) and a CP map $\hat{\Phi}$, we have $\hat{\Phi}(\rho_{S})=\rho_{S}$ for all $\rho_{S}\in \mathcal{S}_{S}$, then there exists a decomposition of the Hilbert space $\cH_S$ as $\cH_S=\bigoplus_{i}\cH_{L_{i}}\otimes\cH_{R_{i}}$ such that: 

(1) each $\rho_{S}\in \mathcal{S}_{S}$ can be decomposed as
\begin{equation}
\label{eq:fifteen}
\begin{aligned}
\rho_{S}=\bigoplus_{i}p_{i}\,\rho_{L_{i}}\otimes\tilde{\omega}_{R_{i}},
\end{aligned}
\end{equation}
where the probability distribution $\lbrace p_{i}\rbrace$ and states $\rho_{L_{i}}\in \mathcal{D}_{L_{i}}$ are dependent on $\rho_{S}$, but the states $\tilde{\omega}_{R_{i}}\in \mathcal{D}_{R_{i}}$ are fixed for all $\rho_{S}$; and

(2) the unitary $V$ in Eq. (\ref{eq:fourteen}) is in the form of
\begin{equation}
\label{eq:sixteen}
\begin{aligned}
V=\bigoplus_{i}I_{L_{i}}\otimes V_{R_{i}EC},
\end{aligned}
\end{equation}
where $I_{L_{i}}$ is the identity operator on $\cH_{L_{i}}$ and $V_{R_{i}EC}$ is a unitary operator on $\cH_{R_{i}}\otimes\cH_{E}\otimes\cH_{C}$.

Combining Eqs. (\ref{eq:twelve}) and (\ref{eq:sixteen}) gives us:
\begin{equation}
\label{eq:seventeen}
\begin{aligned}
\Lambda=\bigoplus_{i}id_{L_{i}}\otimes \Lambda_{R_{i}},
\end{aligned}
\end{equation}
where $id_{L_{i}}$ is the identity map on $\mathcal{L}(\cH_{L_{i}})$ and $\Lambda_{R_{i}}:\mathcal{L}(\cH_{R_{i}})\rightarrow\mathcal{L}(\cH_{R_{i}}\otimes\cH_{E})$ is a CP assignment map on $\mathcal{L}(\cH_{R_{i}})$. Using Eqs. (\ref{eq:fifteen}) and (\ref{eq:seventeen}), we have:
\begin{equation*}
\Lambda(\rho_{S})=\bigoplus_{i}p_{i}\,\rho_{L_{i}}\otimes \Lambda_{R_{i}}(\tilde{\omega}_{R_{i}}),
\end{equation*} 
for each $\rho_{S}\in \mathcal{S}_{S}$. $\Lambda_{R_{i}}$ is a CP map (on $\mathcal{L}(\cH_{R_{i}})$), so it maps the state $\tilde{\omega}_{R_{i}}$ to a state $\tilde{\omega}_{R_{i}E}\in\mathcal{D}_{R_{i}E}$. Therefore:
\begin{equation}
\label{eq:eighteen}
\begin{aligned}
\rho_{SE}=\Lambda(\rho_{S})=\bigoplus_{i}p_{i}\,\rho_{L_{i}}\otimes\tilde{\omega}_{R_{i}E},
\end{aligned}
\end{equation}
which is in the form of Eq. (\ref{eq:ten}). Note that $\tilde{\omega}_{R_{i}}$ are fixed for all $\rho_{S}\in \mathcal{S}_{S}$, and so are $\tilde{\omega}_{R_{i}E}$.

In summary, we have seen that if:

(1) $\mathcal{V}$ is a $U$-consistent subspace of $\mathcal{L}(\cH_{S}\otimes\cH_{E})$ for all $U$, and

(2) (the extension of) the assignment map $\Lambda\,:\mathcal{L}(\cH_{S})\rightarrow\mathcal{L}(\cH_{S}\otimes\cH_{E})$ is CP,\\
then each $\rho_{S}\in \mathcal{S}_{S}=\mathrm{Tr_{E}}(\mathcal{V})\cap\, \mathcal{D}_{S}$ is in the form of Eq. (\ref{eq:fifteen}), and is mapped by the assignment map $\Lambda$ to a $\rho_{SE}\in \mathcal{S}\equiv\mathcal{V}\cap\, \mathcal{D}_{SE}$ which is given by Eq. (\ref{eq:eighteen}). Also note that assuming that the assignment map $\Lambda$ is CP results in the complete positivity of the reduced dynamical map $\Psi=\mathrm{Tr_{E}}\circ Ad_{U}\circ\Lambda$.

In other words, the complete positivity of the assignment map $\Lambda$ assures that the set of initial states $\rho_{SE}$ (leading to CP reduced dynamics) is given by Eq. (\ref{eq:ten}) \cite{16}. Reversely, if the set of initial states $\rho_{SE}$ is given by $\mathcal{S}$ in Eq. (\ref{eq:ten}), then it can be shown simply, by explicit construction of $\Lambda$ \cite{16}, that the assignment map $\Lambda$ is CP.

Let's end this section with the following point.
In the previous section, we have seen that the set  $\mathcal{S}$ in Eq. (\ref{eq:ten}) can be written as the steered set, i.e. as  Eq.  (\ref{eq:1a}), from the fixed Markov state $\tilde{\omega}_{ASE}$, in Eq. (\ref{eq:ten-a}). For the  $U$-consistency of a steered set $\mathcal{S}$, for arbitrary $U\in\mathcal{L}(\cH_{S}\otimes\cH_{E})$, a one-to-one correspondence between the members of $\mathcal{S}$  and $\mathcal{S}_{S}=\mathrm{Tr_{E}}(\mathcal{S})$ is required. For a Markov state $\tilde{\omega}_{ASE}$, the existence of this one-to-one correspondence is proven in Ref. \cite{99}.

\section{When $\mathcal{V}$ is  $U$-consistent for a restricted set of unitary operators $U$ } \label{sec:U-consistent}

In the previous section, we have restated the main result of  Ref.  \cite{16}. We saw that the two conditions of $U$-consistency \textit{for arbitrary }$U$ and the CP-ness of the assignment map, lead to Eq. (\ref{eq:ten}). When the set of initial states $\rho_{SE}$ is given by $\mathcal{S}$ in Eq. (\ref{eq:ten}), then the reduced dynamics , \textit{for arbitrary} $U$, is CP.

 What if we relax the condition of  $U$-consistency for arbitrary $U$? As we will see in the following, this relaxation leads to a generalization of Eq. (\ref{eq:ten}); i.e. we will find a set $\mathcal{S}$ of initial $\rho_{SE}$ which includes the set given in Eq. (\ref{eq:ten}) as a subset and also leads to CP reduced dynamics (for a restricted set of unitary operators $U$).

If $\mathcal{V}$ is  $U$-consistent for all $U$ in a set $\mathcal{G}\subseteq\mathcal{U}(\cH_{S}\otimes\cH_{E})$, then $\mathcal{V}$ is called $\mathcal{G}$-consistent ($\mathcal{U}(\cH_{S}\otimes\cH_{E})$, is the set of all unitary $U\in\mathcal{L}(\cH_{S}\otimes\cH_{E})$). In the previous section, we have seen that if $\mathcal{G} =\mathcal{U}(\cH_{S}\otimes\cH_{E})$, then there is a one-to-one correspondence between the members of $\mathrm{Tr}_{E}(\mathcal{V})$ and $\mathcal{V}$. But if $\mathcal{G}$ is a proper subset of $\mathcal{U}(\cH_{S}\otimes\cH_{E})$, then there is no guarantee for such a one-to-one correspondence. Therefore, one may find two different $X_{1}$ and $X_{2}\in\mathcal{V}$ for which we have $\mathrm{Tr}_{E}X_{1}=\mathrm{Tr}_{E}X_{2}=x$. So, for $Y= X_{1}-X_{2}$ we have $\mathrm{Tr}_{E}Y= 0$. In other words, there is a subset $\mathcal{V}_{0}\subset\mathcal{V}$ such that for each $Y \in\mathcal{V}_{0}$ we have $\mathrm{Tr}_{E}Y= 0$.

Therefore, there is more than one way to define the assignment map. We write:
\begin{equation}
\label{eq:nineteen}
\begin{aligned}
\tilde{\Lambda}=\Lambda+\mathcal{V}_{0},
\end{aligned}
\end{equation}
i.e. if $\Lambda$ is an assignment map from $\mathrm{Tr}_{E}(\mathcal{V})$ to $\mathcal{V}$ such that $\Lambda(x)=X$ ($x\in\mathrm{Tr}_{E}(\mathcal{V})$ and $X\in \mathcal{V}$),  then adding   $Y\in \mathcal{V}_{0}$ gives us another possible assignment map $\tilde{\Lambda}$.

In general, the assignment map $\Lambda$ is Hermitian \cite{3}. But, in this paper, we restrict ourselves to the case that the assignment map $\Lambda$ is, in addition, CP; i.e. $\Lambda$ in Eq. (\ref{eq:nineteen}) is a CP map as Eq. (\ref{eq:four}). Therefore, for this part of the general assignment map $\tilde{\Lambda}$ in Eq. (\ref{eq:nineteen}), a similar line of reasoning, as appeared in the previous section, can be applied. So, for $\Lambda$ in Eq. (\ref{eq:nineteen}), we can write Eq. (\ref{eq:twelve}) and then Eq. (\ref{eq:thirteen}) and finally, using the result of Ref. \cite{17}, achieve Eq. (\ref{eq:eighteen}).

Using Eqs. (\ref{eq:eighteen}) and (\ref{eq:nineteen}), for each $\rho_{S}\in\mathcal{S}_{S}=\mathrm{Tr_{E}}(\mathcal{V})\cap\, \mathcal{D}_{S}$, we have
\begin{equation}
\label{eq:twenty}
\begin{aligned}
\tilde{\Lambda}(\rho_{S})=\Lambda(\rho_{S})+\mathcal{V}_{0}\qquad\qquad \;\;\\
= \bigoplus_{i}p_{i}\,\rho_{L_{i}}\otimes\tilde{\omega}_{R_{i}E} +\mathcal{V}_{0},
\end{aligned}
\end{equation}
where, as before, the probability distribution $\lbrace p_{i}\rbrace$ and states $\rho_{L_{i}}\in \mathcal{D}_{L_{i}}$ are dependent on $\rho_{S}$, but the states $\tilde{\omega}_{R_{i}E}\in \mathcal{D}_{R_{i}E}$ are fixed for all $\rho_{S}\in\mathcal{S}_{S} $. 

Therefore, for the set
\begin{equation}
\label{eq:twenty one}
\begin{aligned}
\mathcal{S}\equiv\mathcal{V}\cap\, \mathcal{D}_{SE}=\lbrace 
 (\bigoplus_{i}p_{i}\,\rho_{L_{i}}\otimes\tilde{\omega}_{R_{i}E} +\mathcal{V}_{0})\cap\, \mathcal{D}_{SE} \rbrace, 
\end{aligned}
\end{equation}
the reduced dynamical map $\Psi=\mathrm{Tr_{E}}\circ Ad_{U}\circ\tilde{\Lambda}$ is CP, since the assignment map $\Lambda$ is CP. Obviously, Eq. (\ref{eq:ten}) is a special case of Eq. (\ref{eq:twenty one}), due to the case that $\mathcal{V}_{0}$ is null, which is for $\mathcal{G}=\mathcal{U}(\cH_{S}\otimes\cH_{E})$.

In summary, we have proved the following theorem:

\textbf{Theorem 1.}  \textit{If}

\textit{(1)  $\mathcal{V}$ is a $U$-consistent subspace of $\mathcal{L}(\cH_{S}\otimes\cH_{E})$, for all $U\in \mathcal{G}$ where $\mathcal{G}\subseteq\mathcal{U}(\cH_{S}\otimes\cH_{E})$  , and}

\textit{(2)  (the extension of) the assignment map $\Lambda\,:\mathcal{L}(\cH_{S})\rightarrow\mathcal{L}(\cH_{S}\otimes\cH_{E})$ is CP,\\
then $\mathcal{S}=\mathcal{V}\cap\, \mathcal{D}_{SE}$ is given by Eq. (\ref{eq:twenty one}), which leads to CP reduced dynamical maps $\Psi=\mathrm{Tr_{E}}\circ Ad_{U}\circ\tilde{\Lambda}=\mathrm{Tr_{E}}\circ Ad_{U}\circ\Lambda$ for all $U\in \mathcal{G}$.}

Since  Eq. (\ref{eq:ten}) is a special case of Eq. (\ref{eq:twenty one}), the above theorem includes all the previous results in this context as special cases. 
We end this section with two simple illustrating examples.

\textbf{Example 1.} Consider the case that $\mathcal{S}=\lbrace \rho_{SE}\in \mathcal{D}_{SE} : \mathrm{Tr_{S}}(\rho_{SE})=\tilde{\omega}_{E}\rbrace$, where $\tilde{\omega}_{E}\in \mathcal{D}_{E}$ is a fixed state, and $\mathcal{G}=\lbrace U_{sw}\rbrace$, where $U_{sw}$ is the swap operator. $\mathcal{S}$ is convex and $\mathcal{G}$-consistent.

Equivalently, we can write $\mathcal{S}$ as 
\begin{equation}
\label{eq:twenty two}
\begin{aligned}
\mathcal{S}=\lbrace (\rho_{S}\otimes\tilde{\omega}_{E} + \mathcal{V}_{0}) \cap\, \mathcal{D}_{SE} \rbrace,
\end{aligned}
\end{equation}
 where $\rho_{S}$ is an arbitrary state in  $\mathcal{D}_{S}$. For this example, the assignment map $\Lambda$ in Eq. (\ref{eq:nineteen}) is CP, since, for each $x\in\mathrm{Tr}_{E}(\mathcal{V})=\mathcal{L}(\cH_{S})$, we have $\Lambda (x)=x\otimes\tilde{\omega}_{E}$ which is obviously a CP map.
 
 Therefore, for the set $\mathcal{S}$ in  Eq. (\ref{eq:twenty two}), the reduced dynamical map $\Psi=\mathrm{Tr_{E}}\circ Ad_{U_{sw}}\circ\tilde{\Lambda}$ is CP. The set $\mathcal{S}$ in  Eq. (\ref{eq:twenty two}) obviously includes Eq. (\ref{eq:five-a}) as a subset. Recall that Eq. (\ref{eq:five-a}) is the set given by Eq. (\ref{eq:ten}) for the case that the summation in Eq. (\ref{eq:ten}) includes only one term and $\cH_{R}$ is a trivial one-dimensional Hilbert space. 
 
 \textbf{Example 2.} Consider the case that $\mathcal{S}=\mathcal{D}_{SE}$ and $\mathcal{G}=\lbrace U_{S}\otimes U_{E}\rbrace$, where $U_{S}$ ($U_{E}$) are arbitrary unitary operators on $\cH_{S}$ ($\cH_{E}$). 
$\mathcal{S}$ is convex and  $\mathcal{G}$-consistent and we can write it as
\begin{equation}
\label{eq:twenty three}
\begin{aligned}
\mathcal{S}=\lbrace (\rho_{S}\otimes\tilde{\omega}_{E} + \mathcal{V}_{0}) \cap\, \mathcal{D}_{SE} \rbrace,
\end{aligned}
\end{equation}
where $\rho_{S}$ is an arbitrary state in  $\mathcal{D}_{S}$ and  $\tilde{\omega}_{E}$ is (arbitrary chosen) fixed state in  $\mathcal{D}_{E}$. Note that $\mathcal{S}$ in  Eq. (\ref{eq:twenty three}) differs from $\mathcal{S}$ in  Eq. (\ref{eq:twenty two}) , because of the difference between $\mathcal{V}_{0}$ in the two mentioned equations. In the current example, $\mathcal{V}=\mathcal{L}(\cH_{S}\otimes\cH_{E})$. So $\mathcal{V}_{0}= \mathrm{ker}\ \mathrm{Tr_{E}}$, i.e. the set of all $Y \in \mathcal{L}(\cH_{S}\otimes\cH_{E})$ for which $\mathrm{Tr_{E}}(Y)=0$.

Writing $\mathcal{S}$ as  Eq. (\ref{eq:twenty three}) shows that the assignment map $\Lambda $ in this example is the same as $\Lambda $ in the previous example. The CP-ness of $\Lambda $ results in the CP-ness of  the reduced dynamical maps $\Psi=\mathrm{Tr_{E}}\circ Ad_{U}\circ\tilde{\Lambda}$, for all $U \in \mathcal{G}$.

Obviously, the set $\mathcal{S}=\mathcal{D}_{SE}$ in this example, which yields CP reduced dynamics, is larger than any set which can be constructed by Eq. (\ref{eq:ten}).

Note that this example is, in fact, restating the result of Ref. \cite{6}, using the  Theorem 1.

It is also worth noting that though the assignment map  $\Lambda $, in the above two examples, became the same as the one first introduced by Pechukas \cite{21,8}, but, because in the above examples $\mathcal{G}\neq \mathcal{U}(\cH_{S}\otimes\cH_{E})$  and so $\mathcal{V}_{0}$ is not null, the assignment map  $\tilde{\Lambda} $ is not the same as the Pechukas's one.

\section{Summary}\label{sec:summary}

Though there was a tendency to assume the CP maps as the only possible quantum dynamics of a system, we have seen in Sec.~\ref{sec:reduced_dynamics} that this is not the case, at least, for open quantum systems.

In fact, the CP-ness of the reduced dynamics of the system, for arbitrary $U$ in Eq. (\ref{eq:two}), has been proven only for some restricted sets of initial $\rho_{SE}$, which we reviewed them in Sec.~\ref{sec:CP reduced_dynamics}. All the sets given in this section can be written as special cases of Eq. (\ref{eq:ten}). 

An important result in this context is that of Ref. \cite{16}, where it has been shown that if the assignment map is CP, then the set of initial $\rho_{SE}$ is given by Eq. (\ref{eq:ten}).

In Sec.~\ref{sec:assignment map}, we restated the mentioned result of Ref. \cite{16}, using the framework introduced in Ref. \cite{3}. This treatment highlighted the condition of  $U$-consistency for arbitrary $U$, which is needed to achieve the result of Ref. \cite{16}. In Ref. \cite{16}, it has been assumed implicitly
 that there is only one way to define the assignment map, which is true only when the $U$-consistency condition is valid for arbitrary $U$. 

Finally, our treatment  led us to the main result of this paper, which was given in Sec.~\ref{sec:U-consistent}. There, we relaxed the condition of $U$-consistency for arbitrary $U$ and found the most general possible set of initial $\rho_{SE}$ for which the assignment map $\Lambda $ is CP and so the reduced dynamical maps are also CP (for a restricted set of $U$ in Eq. (\ref{eq:two})). This set, which was given in  Eq. (\ref{eq:twenty one}), includes  Eq. (\ref{eq:ten}) as a special case. Therefore, it includes all the previously found sets as special cases as well.


\end{document}